# Impact of Electrostatic Doping Level on the Dissipative Transport in Graphene Nanoribbons Tunnel Field-Effect Transistors

Weixiang Zhang[1], Tarek Ragab[2]*, Ji Zhang[1], Cemal Basaran[1]

[1]Electronic Packaging Laboratory, Buffalo, NY 14260

[2]Arkansas State University, State University, AR 72467

*Corresponding author: tragab@astate.edu

## Abstract

The impact of electrostatic doping level on the dissipative transport of Armchair GNR-TFET is studied using the Quantum Perturbation Theory (QPT) with the Extended Lowest Order Expansion (XLOE) implementation method. Results show that the doping level of the source and drain sides of the GNR-TFET has a significant impact on the phonon contribution to the carrier transport process. Unlike in other similar studies, where phonons are believed to have a constant detrimental influence on the $I_{ON}/I_{OFF}$ ratio and Subthreshold Swing (SS) of the TFET devices due to the phonon absorption-assisted tunneling, we show that by a proper engineering of the doping level in the source and drain, the phonon absorption assisted tunneling can be effectively inhibited. We also show that as temperature increase, the device switching property deteriorates in both the ballistic and dissipative transport regimes, and there exists a temperature-dependent critical doping level where the device has optimal switching behavior.

## Introduction

Tunneling field effect transistor (TFET) is considered as a promising candidate to substitute conventional metal-oxide semiconductor field-effect transistor (MOSFET for energy efficient applications due to their small OFF state current and reducible Subthreshold Swing (SS) that can operate below the conventional MOSFET limit of 60 mv/decade at room temperature [1-3].

Typically, graphene nanoribbons (GNR) are one of the most intensively studied material system for TFET applications due to its exceptional mechanical, thermal and electrical properties [4-9]. In order to accurately model the carrier transport in GNR-TFET device, it is important to account for the electron-phonon interactions. Phonons could contribute significantly to the Band-To-Band Tunneling (BTBT) and the dissipative transport in GNR-TFET can be fundamentally different from the ballistic transport regime [10].

A few groups have studied the role of electron-phonon interaction and the electron inelastic scattering using various models. Yoon et al. [11] calculated the device characteristics for 16 dimer armchair GNR TFET using self-consistent 3D Non-Equilibrium Green's Function (NEGF) with the tight-binding approximation. They show a transition from ballistic to dissipative transport at varying channel length. They also show that the acoustic phonon (AP) scattering is the dominant scattering mechanism within the relevant bias range. Grassi et al. [12] studied the phonon scattering in GNR-TFET using NEGF with the mode-space approach. They showed that the Optical Phonon (OP) scattering significantly increases the minimum leakage current and the sub-threshold slope due to phonon-assisted BTBT, while the AP scattering has a negligible effect on the OFF state current but would reduce the ON state current. Salahuddin et al. [13] studied the dissipative transport in armchair GNR-TFET using NEGF with the real-space approach. They show that the dissipative scattering imposes a limit on the OFF state current and SS for device channel length greater than 15nm due to OP assisted tunneling.

These studies reveal certain aspects of the nature of dissipative transport in GNR-TFET. However, to get a better understanding, further explorations are still needed. In this paper, the important effect of doping level on the dissipative carrier transport in GNR-TFET is explored. Using quantum perturbation theory with the Extended Lowest Order Expansion (XLOE) implementation, we simulate both the ballistic and dissipative transfer characteristics of 7-dimer armchair GNR-TFET at different source and drain doping levels. In addition, the temperature effect on the transfer characteristics is also studied.

In the following sections, the device configuration and the simulation method is presented, then the transfer characteristics with/without phonon effect is derived for the 7-dimer armchair GNR-TFET at three different doping levels. The Projected Local Density of States (PLDOS), spectral current and the phonon dispersion relation are analyzed. Finally the doping-level effect at different temperatures is explored.

**Device parameters and simulation methodology**

The schematic device configuration of the GNR-TFET, as shown in Fig.1, consists of the source, channel, drain region and the two electrodes, with the channel length $L_C$ = 16nm (shorter than sub-10 nm GNRs mean free path ~10 nm [14]), and the source/drain length $L_{S/D}$ = 6nm. Hydrogen-passivated homogenous 7-dimer armchair GNR is used, which has a width of $W_{GNR}$ = 0.8 nm and an energy bandgap $E_g$ = 0.96 eV and is the narrowest GNR that is available experimentally [15]. In order to enhance the electrostatic gate control, the double-sided gate is utilized and the 7-dimmer armchair GNR is sandwiched between two diamond dielectric layers ($\varepsilon_r$ = 5.7) with a thickness $T_d$ = 1 nm. The source and drain regions are electrically doped to be p- and n-type, respectively, by the two doping gates on two sides, while the gate in the middle is the regular control gate that governs the opening and closing of the channel.

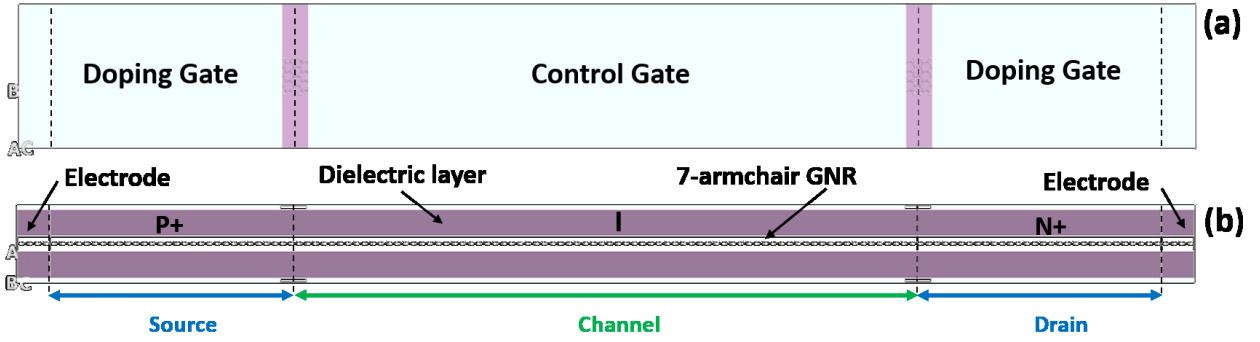

Fig.1 (a) top view and (b) side view of the schematic GNR TFET configuration.

A common supply voltage of 0.2 V [16] is used in all simulations and to explore the temperature effect, the temperature T is chosen to be 300 K to 1200K with increments of 300K. In order to investigate the doping level effect on the dissipative carrier transport, three doping gate voltages are chosen, i.e. ± 1.5 V, ± 2.3 V, and ± 3.5 V for source and drain, respectively.

The dissipative carrier transport is simulated using the Extended Lowest Order Expansion method (XLOE) [17-21], which is an approximation method that circumvents the Self-consistent Born approximation (SCBA) calculation by expand it to the second order to avoid direct integral. The detailed steps for calculating the dissipative transport are described as follows. First, the dynamical matrix and Hamiltonian derivatives of the device configuration need to be calculated by the following equations:

$$D_{\mu\alpha,i\beta} = \frac{dF_{i\beta}}{dr_{\mu\alpha}} \approx \frac{F_{i\beta}(\Delta r_\alpha) - F_{i\beta}(-\Delta r_\alpha)}{2\Delta r_\alpha} \quad (1)$$

and

$$\langle i| \frac{\partial \widehat{H}}{\partial R_{I,\alpha}} |j\rangle \approx \frac{\partial}{\partial R_{I,\alpha}} \big(\langle i|\widehat{H}|j\rangle\big) \approx \frac{H_{ij}(R_{I,\alpha}+\delta) - H_{ij}(R_{I,\alpha}-\delta)}{2\delta} \quad (2)$$

where $D_{\mu\alpha,i\beta}$ denote the dynamic matrix element in Cartesian direction α and β directions (i.e. *x, y, z*), $F_{i\beta}$ is the force on atom *i* in the direction β due to a displacement of atom μ in direction α. $\hat{H}$ is the device Hamiltonian and $R_{I,\alpha}$ is the α Cartesian coordinate for atom *I* in the central unit cell, $\delta$ is the atomic displacement. Next, the phonon frequency $\omega_\lambda$ and eigenvectors $\mathbf{u}^\lambda$ can be obtained by:

$$D\mathbf{u}^\lambda = \omega_\lambda^2 \mathbf{u}^\lambda \quad (3)$$

Next, one can obtain the electron-phonon coupling matrix for given phonon mode λ by:

$$M_{ij}^\lambda = \sum_{Iv} \langle i| \frac{\partial \hat{H}}{\partial R_{Iv}} |j\rangle \mathbf{v}_{Iv} \quad (4)$$

where the $\mathbf{v}_{Iv}$ is defined as:

$$\mathbf{v}_{Iv} = \mathbf{u}_{Iv}^\lambda \sqrt{\frac{\hbar}{2M_I \omega_\lambda}} \quad (5)$$

and *I* is atom indices and *v* is the Cartesian direction (*x, y, z*). After calculating the electron-phonon coupling matrix, the symmetric inelastic transmission functions $T_\lambda^{sym}$ can be obtained by:

$$T_\lambda^{sym} = Tr\left[G^r \Gamma_L G^a \left\{ M^\lambda A_R M^\lambda + \frac{i}{2}(\Gamma_R G^a M^\lambda (A_R + A_L) M^\lambda) \right\} \right] \quad (6)$$

where $G^r$ ($G^a$) is the retard (advanced) Green's functions [22], $\Gamma_{L/R} = i(\Sigma_{L/R}^* - \Sigma_{L/R})$ is the broadening function, $\Sigma_{L/R}$ is the self-energies of the left/right electrode which describe the effect of the electrode states on the electronic structure of the channel region and is calculated using a iterative scheme [22,23], $A_{L/R} = G^r \Gamma_{L/R} G^a$ is the spectral function. Finally, the dissipative current ca be calculated as [24]:

$$I(V) = I_0(V) + \sum_\lambda \int_{-\infty}^{\infty} d\epsilon [T_\lambda^{sym,pos}(\epsilon) F_{pos}^\lambda(\epsilon) + T_\lambda^{sym,neg}(\epsilon) F_{neg}^\lambda(\epsilon)] \quad (7)$$

where $I_0(V)$ is the ballistic current calculated using Landauer-Bütiker formula [5], the $T_\lambda^{sym,pos/neg}$ is symmetric inelastic transmission at positive and negative bias, and $F_{pos/neg}^\lambda$ is the energy and mode depended pre-factors:

$$F_{pos}^\lambda = n_F(\mu_L)[1 - n_F(\mu_R - \hbar\omega_\lambda)][n_B(\hbar\omega_\lambda) + 1] - n_F(\mu_R + \hbar\omega_\lambda)[1 - n_F(\mu_L)]n_B(\hbar\omega_\lambda) \quad (8)$$

and

$$F_{neg}^\lambda = n_F(\mu_L)[1 - n_F(\mu_R + \hbar\omega_\lambda)]n_B(\hbar\omega_\lambda) - n_F(\mu_R - \hbar\omega_\lambda)[1 - n_F(\mu_L)][n_B(\hbar\omega_\lambda) + 1] \quad (9)$$

where $n_F$ and $n_B$ are the Fermi-Dirac and Bose-Einstein distribution functions.

## Results and Discussion

In order to investigate the doping level effect on the dissipative carrier transport at room temperature, three doping gate voltage values, i.e. ± 1.5 V, ± 2.3 V and ± 3.5 V, are examined and their corresponding transfer characteristics are obtained as shown in Fig.2

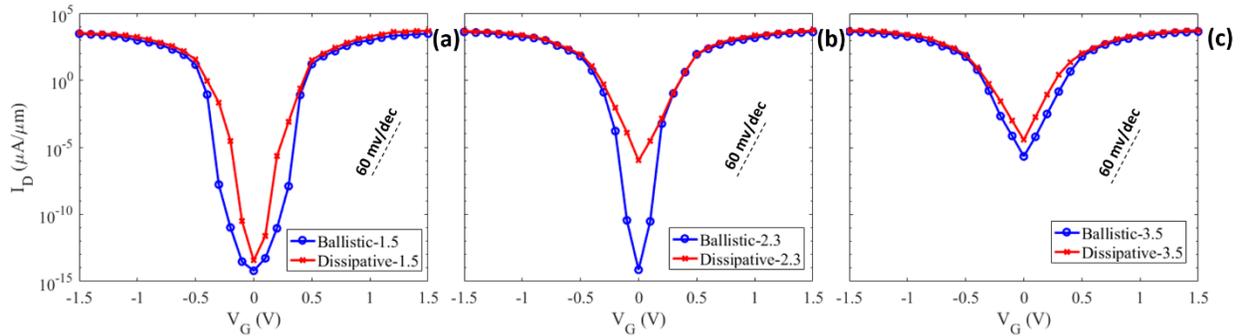

Fig.2. Ballistic and dissipative transfer characteristics of 7-dimer armchair GNR-TFET with (a) doping voltage ± 1.5 V, (b) doping voltage ± 2.3 V and (c) doping voltage ± 3.5 V at source/drain, respectively, at 300 K.

It can be seen that the ON state current at all three doping voltages have a comparable magnitude with both ballistic and dissipative assumptions, while the OFF state current is significantly different at different doping voltages. The difference between ballistic current $I_{bal}$ and dissipative current $I_{diss}$ at OFF state is relatively small in the ± 1.5 V (Fig.2a) and ± 3.5 V (Fig.2c) doping voltages, while for the ± 2.3V case (Fig.2b), the OFF state $I_{diss}$ is 8 orders of magnitude larger than that of $I_{bal}$, indicating a strong phonon contribution on the carrier transport. The difference between OFF state $I_{bal}$ and $I_{diss}$ can be further explained by analyzing the Projected Local Density of States (PLDOS), the spectral current shown in Fig.3 and the 7-dimer armchair GNR phonon dispersion relation shown in Fig.4.

For the ± 1.5V case (Fig.3a and Fig.3d), the energy difference ΔE between $E_{SV}$ (Source side Valence band) and $E_{CC}$ (Channel Conduction band) as well as $E_{CV}$ (Channel Valence band) and $E_{DC}$ (Drain side Conduction band) is 320 meV, which is beyond the range of the maximum phonon energy of the 7-dimer armchair GNR of 220meV as shown in Fig.4. Thus the electrons that reside at energy level $E_{SV}$/$E_{CV}$ cannot gain enough energy to jump into the energy states that are higher than $E_{CC}$/$E_{DC}$ through phonon absorption. As a result, the phonon-assisted band-to-band tunneling (BTBT) leakage path cannot be established at the OFF state, and the majority of the spectral current is distributed within the bias window [-0.1, 0.1] eV and its vicinity for both the ballistic and dissipative transport. Due to the presence of phonon scattering, the range of direct source-to-drain tunneling is broadened, and the OFF state $I_{diss}$ is relatively larger compared to the OFF state $I_{bal}$ as shown in part d of figure 3.

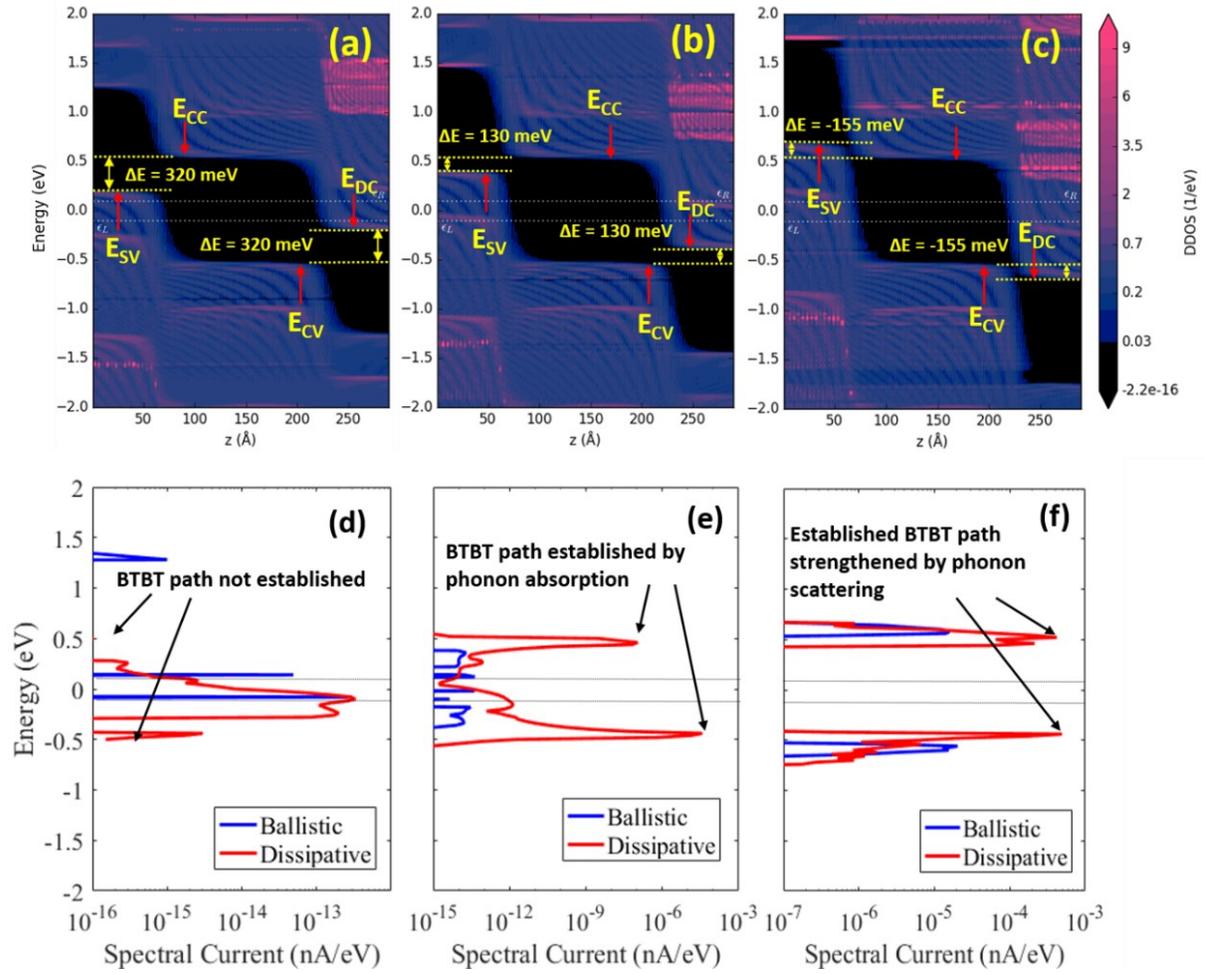

Fig.3 Spatially resolved (in the transport direction) Projected Local Density of States for OFF state 7-dimer armchair GNR-TFET at different doping voltage (a) ± 1.5V, (b) ± 2.3V and (c) ± 3.5V, and the corresponding ballistic and dissipative spectral current in semi-log scale for (d) ± 1.5V, (e) ±2.3V, and (f) ±3.5V.

When the control gate voltage $V_G$ shifts away from 0V to the subthreshold region (i.e. [-0.5V, 0V] or [0V, 0.5V]) the $E_{CC}$ and $E_{CV}$ move up/down in energy position, and the energy difference ΔE between $E_{CV}/E_{SV}$ and $E_{DC}/E_{CC}$ decreases. As a result, phonon-assisted BTBT path can be created in subthreshold region and the dissipative current $I_{diss}$ increases faster than the ballistic current $I_{bal}$. Therefore, a smaller sub-threshold swing (SS) can be observed for the dissipative branch.

For the ±2.3V case (Fig.3b and Fig.3e), the energy difference ΔE between $E_{SV}/E_{CV}$ and $E_{CC}/E_{DC}$ is 130 meV, which is smaller than the maximum phonon energy of 7-dimer armchair GNR. Thus phonon-assisted BTBT leakage paths can be created and two spectral current peaks (the red dotted lines of figure 3-e) can be observed at ~ ± 0.5eV. Since the barrier distance in phonon-assisted BTBT is much smaller than the direct source-to-drain tunneling distance (the channel length), the tunneling probability of the phonon-assisted BTBT is much larger and the OFF state $I_{diss}$ is significantly greater than the OFF state $I_{bal}$, giving rise to a smaller $I_{ON}/I_{OFF}$ ratio and larger SS. Thus, it can be inferred that phonon has a detrimental influence on the 7-dimer armchair GNR device performance at ± 2.3V doping voltage.

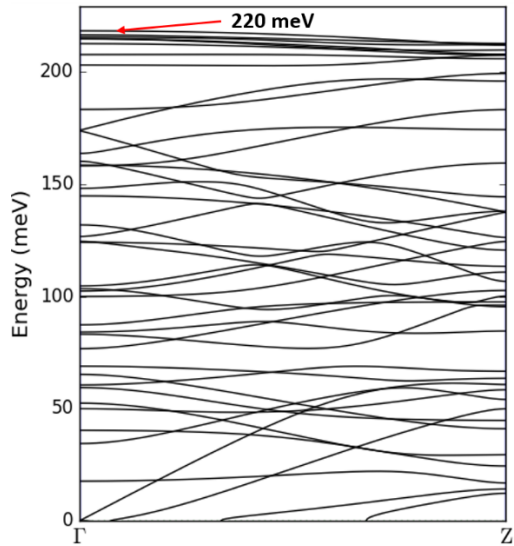

Fig.4 Phonon dispersion relation for 7-dimer armchair GNR, calculated using Tersoff_CH_2010 force field potential [25, 26]. The energy range of all phonon mode is within 0-220 meV.

For the ± 3.5V case (Fig.3c and Fig.3f), since the energy difference ΔE between $E_{SV}/E_{CV}$ and $E_{CC}/E_{DC}$ is negative, i.e. $E_{SV}/E_{CV}$ is larger than $E_{CC}/E_{DC}$, the BTBT paths are already established in the ballistic transport case. Thus, the major current contribution is from the energy position of BTBT ~ ±0.5V as shown in Fig.3f, and the OFF state $I_{bal}$ is significantly larger than the ± 1.5V and ± 2.3V doping voltages. With the presence of phonon scattering, the established BTBT paths are strengthened and broadened in energy range and two larger spectral current peaks can be observed (red dotted lines) in Fig.3f. Unlike the ±2.3V case, the presence of phonon in ± 3.5V case only serves to strengthen the already established BTBT path, thus the difference between OFF state $I_{diss}$ and OFF state $I_{bal}$ is relatively smaller. However, since the leakage BTBT paths are established in OFF state, the dissipative $I_{ON}/I_{OFF}$ ratio for ± 3.5V case is much smaller than that of the ± 1.5V and ± 2.3V cases. Therefore, in designing GNR-TFET with strong switching property, an excessively high doping level at source and drain beyond 1.5 V should be avoided. This is also verified in designing other 2D semiconductor TFET models [27].

Summarizing the findings so far, it is shown that at 300K, the doping level at source and drain has a profound influence on the carrier transport properties. In some doping level, the energy of electrons is elevated due to phonon absorption and parasitic BTBT leakage paths are created at OFF state, which is harmful to the device switching property. By properly engineering the doping level such that the $E_{SV}/E_{CV}$ and $E_{CC}/E_{DC}$ energy difference $\Delta E$ is greater than the maximum phonon energy, the parasitic BTBT paths can be avoided at OFF state and the device switching performance can be retained.

Next, the effect of temperature is studied by comparing the ballistic and dissipative transfer properties attemperatures ranging from 300K to 1200K. As shown in Fig.5, the ON state current in all cases have a comparable magnitude of $10^4$ µA/µm. This is because the major contribution to the ON state current is the direct BTBT, which is relatively insensitive to temperature and doping voltage on the source and the drain. The OFF state current, on the other hand, is largely dependent on the temperature in both ballistic and dissipative transport. Typically, when the temperature increases, the OFF state current increases exponentially, causing the $I_{ON}/I_{OFF}$ ratio to decrease significantly.

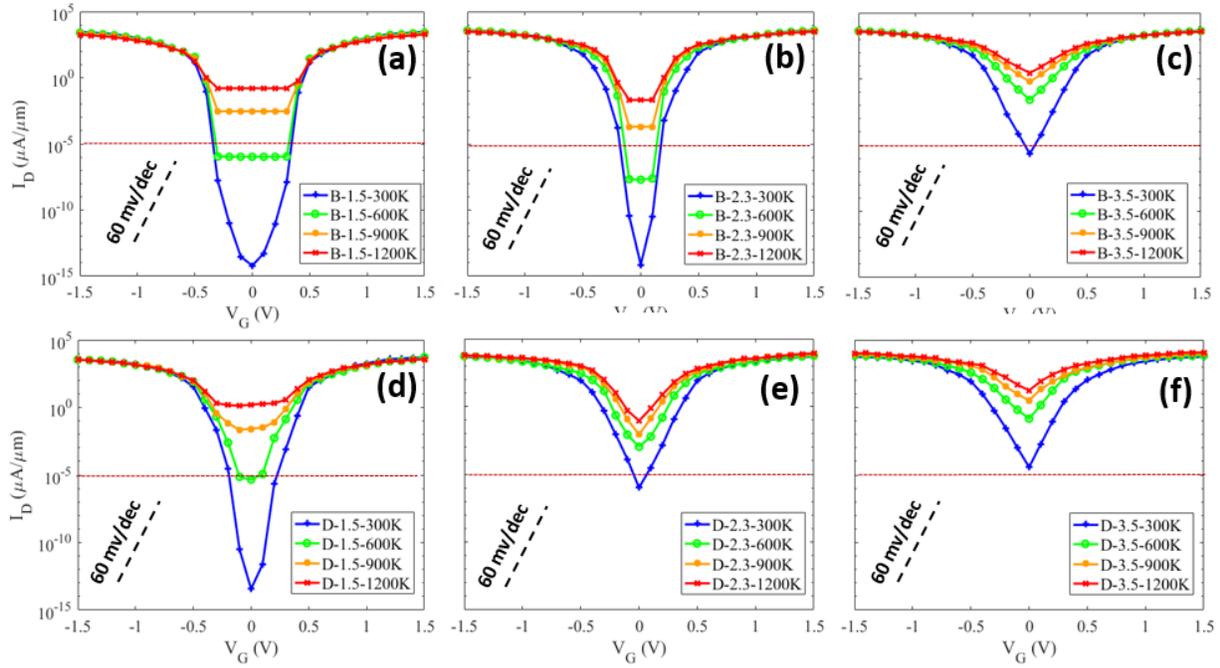

Fig.5 Ballistic and dissipative transport $I_D$-$V_G$ characteristics of 7-dimer armchair GNR-TFET at 300K-1200K and different doping voltages: (a) Ballistic transport with ± 1.5 V doping voltage, (b) Ballistic transport with ± 2.3 V doping voltage, (c) Ballistic transport with ± 3.5 V doping voltage, (d) Dissipative transport ± 1.5 V doping voltage, (e) Dissipative transport ± 2.3 V doping voltage and (f) Dissipative transport ± 3.5 V doping voltage. The red dotted line at $10^{-5}$ µA/µm represents the low power OFF state current level required by ITFS standard [27].

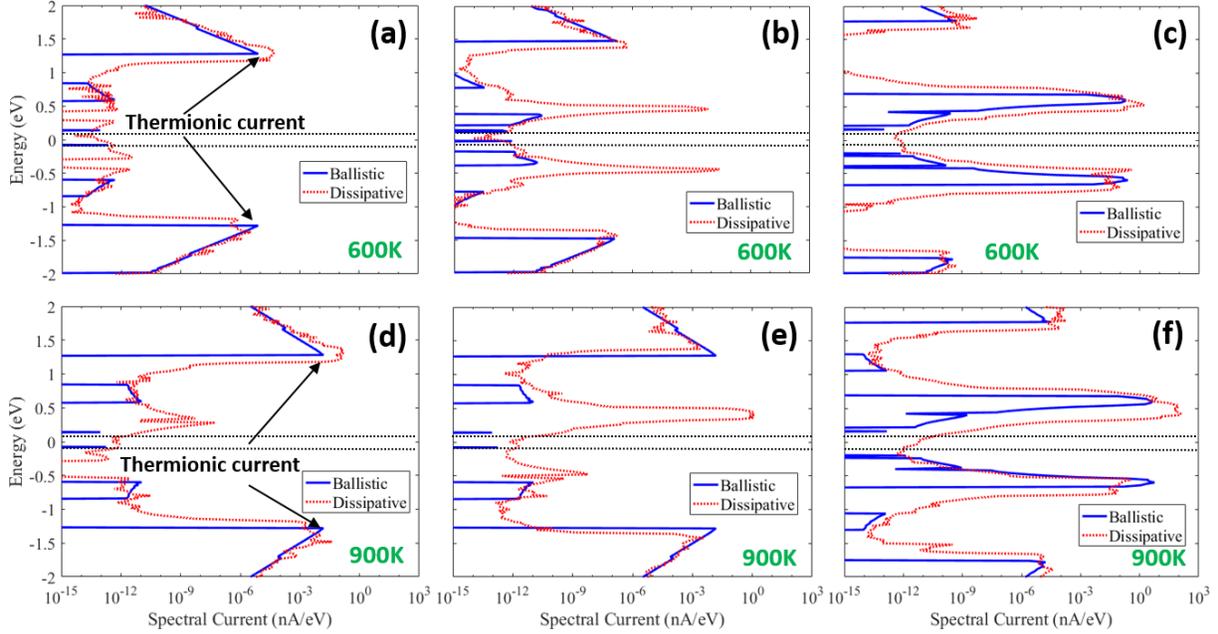

Fig.6 OFF state Spectral current at different temperatures and doping voltages. (a)-(c) 600K, (a) ± 1.5 V (b) ± 2.3 V and (c) ± 3.5 V. (d)-(f) 900K, (d) ± 1.5 V (e) ± 2.3 V and (f) ± 3.5 V. Dotted lines in the middle represent the bias window -0.1eV to 0.1eV.

When the temperature is beyond 600 K, the OFF state $I_{diss}$ is greater than $10^{-5}$ µA/µm in all cases, which is the low power OFF state current level required by the ITRS standard [27]. Thus, the device is no longer considered energy-efficient at elevated temperatures.

It is interesting to note that in the ballistic transport of ± 1.5 V and ± 2.3 V cases, flat bottoms of the $I_D$-$V_G$ are observed for high temperature beyond 300 K (Fig.5a and Fig.5b). This can be explained by analyzing the OFF state spectral current shown in Fig.6. Due to the less steepness of the Fermi-Dirac distribution at elevated temperatures, the electron occupation probability at the tail end of the distribution function are significantly higher. As a result, the thermionic current at the far ends of energy spectrum (Fig.6a, Fig.6b, Fig.6d and Fig.6e) become dominant over the direct source-to-drain tunneling current in the bias window and its vicinity. Thus, the gate voltage $V_G$ shifting would have little effect on increasing the current until $V_G$ shifts far enough from 0V and the direct BTBT could take over again. It is noted that the thermionic spectral current peak in the ± 1.5 V case has a larger magnitude than that of the ± 2.3 V case since it is closer to the Fermi level. As a result, in the ± 1.5 V case the flat bottom is wider and the OFF state $I_{bal}$ is larger compared to that in the ± 2.3V case (Fig.5a and Fig.5b).

On the other hand, since the thermionic spectral current peak also exists in dissipative transport, in the ± 1.5 V case the OFF state $I_{diss}$ at 600K is much greater compared to that of 300K where the thermionic current is trivial. The OFF state $I_{diss}$ increases even more when temperature increases to 900K that the OFF state $I_{diss}$ in the ± 1.5 V case is even larger than that in the ± 2.3 V case (Fig.5d and Fig.5e). Therefore, due to the relatively low doping voltage of ± 1.5 V, the device switching property deteriorate quickly at elevated temperatures.

In order to explore the best switching performance of the 7-dimer armchair GNR-TFET, the OFF state $I_{bal}$ and $I_{diss}$ are plotted against the doping voltage level for 300 K to 900 K, as shown in Fig.7. Note that the source and drain are oppositely doped with an equal amount in all simulations, and the doping gate voltage in Fig.7 represents the absolute value of the source/drain doping. Since the source/drain should be adequately doped to ensure high ON state current, and excessively high doping level deteriorates the switching performance, therefore ± 1.5 V and ± 2.3 V are chosen to be the lower bound and upper bound for the doping voltage range.

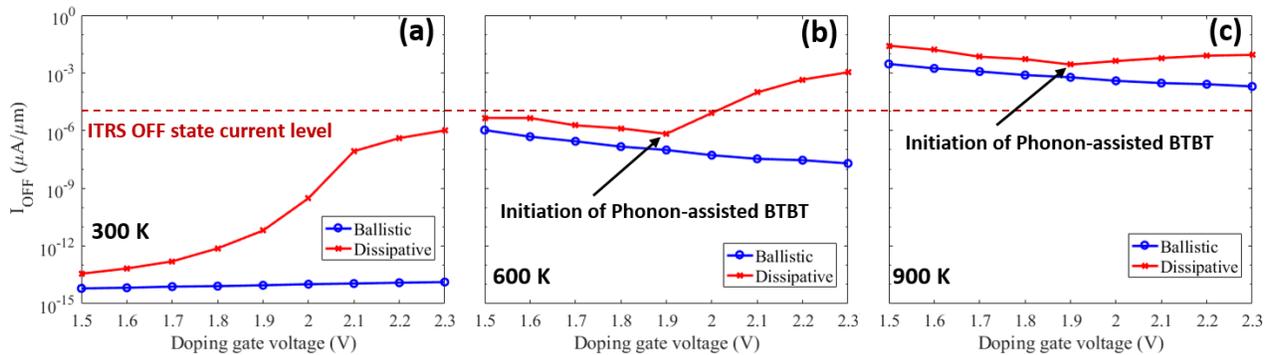

Fig.7 OFF state ballistic and dissipative current versus doping voltage level at 300 K - 900 K. The source and drain are oppositely doped with the same amount, and the doping gate voltage represents the absolute value of their doping.

It can be seen that at 300K (Fig.7a), the OFF state $I_{bal}$ is relatively insensitive to the doping voltage change, while the OFF state $I_{diss}$ increases monotonically with the doping voltage. Thus, it can be easily identified that ± 1.5 V is the critical doping level for optimal switching performance.

For 600K (Fig.7b), the OFF state $I_{bal}$ is monotonically decreasing from ±1.5 V to ±2.3V. This is consistent with our previous analysis that as the doping voltage increases, the intensity of OFF state thermionic spectral current decreases due to its larger distance from the Fermi level and lower occupation probability. On the other hand, the OFF state $I_{diss}$ first decreases, until it reaches the critical point of ±1.9 V where the phonon-assisted BTBT starts and becomes more dominant at larger doping voltage. As a result, the $I_{diss}$ starts to increase with doping voltage after ±1.9 V. Since the OFF state $I_{diss}$ is smallest at ± 1.9 V, it is the critical doping level for optimal device switching performance.

The situation for 900 k is similar to that of 600 K and ± 1.9 V is identified as the critical doping level for optimal switching performance. However, because of the increased occupation at the tail ends of the Fermi-Dirac distribution at high temperature, the OFF state $I_{bal}$ and $I_{diss}$ increase significantly in 900 K. As a result, they are both higher than the ITRS OFF state current level in all doping voltages. Thus, at 900 K the device has a severe leakage problem. At temperature of 1200 K, the device performance is worse and results were not included.

## Conclusions

It is shown that the doping level at source/drain has a profound influence on the dissipative transport of 7-dimer armchair GNR-TFET. Phonon-assisted BTBT path is established at OFF state when ΔE (i.e. the energy difference between the channel conduction band and the source valence band or drain conduction band and channel valence band) is smaller than the maximum phonon energy, and as a result, the $I_{ON}/I_{OFF}$ ratio is greatly compromised compared to its ballistic transport case. By means of engineering the doping voltage such that the ΔE is larger than the maximum phonon energy, the detrimental phonon effect can by largely inhibited and the device switching performance can be retained. Meanwhile, it is shown that the device switching performance deteriorates at high temperatures and the 7-dimer armchair GNR-TFET functions properly up to 600 K. For temperatures beyond 600 K, the device is failing because of severe leakage current at OFF state. In addition, we show that by fine-tuning the source/drain doping voltage and keeping them at the critical doping level, the device switching performance can be kept optimal at different temperatures.

## Acknowledgment

This project is sponsored by US Navy, Office of Naval Research, Advanced Naval Research Platform Power and Energy Program (Grant Number N00014-151-2216) under the direction of Program Director Capt. Lynn Petersen.